\documentclass[10pt, journal]{IEEEtran}

\usepackage{graphicx, enumerate}
\usepackage{cite,array, bm}
\usepackage{dsfont, epstopdf}
\usepackage{mathrsfs, subfigure, color}
\usepackage{amssymb, multirow}
\usepackage[cmex10]{amsmath}
\usepackage{subeqnarray, tabularx}
\usepackage{cases, stmaryrd}
\usepackage{algpseudocode}
\usepackage[ruled,vlined,linesnumbered]{algorithm2e}
\usepackage[utf8]{inputenc}
\usepackage{threeparttable}
\usepackage{stackengine}
\usepackage{verbatim}

\usepackage{enumitem}
\newlist{myenumi}{description}{10}
\setlist[myenumi]{labelindent=\parindent, leftmargin=*, label=(\roman*), align=left}
\setlist[myenumi]{leftmargin=5pt}

\DeclareFontFamily{U}{FdSymbolC}{}
\DeclareFontShape{U}{FdSymbolC}{m}{n}{<-> s * FdSymbolC-Book}{}
\DeclareSymbolFont{fdarrows}{U}{FdSymbolC}{m}{n}
\DeclareMathSymbol{\vDdash}{\mathrel}{fdarrows}{254}

\DeclareFontFamily{U}{FdSymbolD}{}
\DeclareFontShape{U}{FdSymbolD}{m}{n}{<-> s * FdSymbolD-Book}{}
\DeclareSymbolFont{fdnarrows}{U}{FdSymbolD}{m}{n}
\DeclareMathSymbol{\nvDdash}{\mathrel}{fdnarrows}{254}
\makeatother

\newtheorem{theorem}{Theorem}

\newtheorem{proposition}{Proposition}
\newtheorem{problem}{Problem}

\newcommand{\vol}{\textsf{vol}}
\newcommand{\evl}{\textsf{evl}}

\definecolor{bluencs}{rgb}{0.0, 0.53, 0.74}

\begin{document}

\title{\LARGE \bf Reachability-based Control Synthesis under Signal Temporal \\ Logic Specifications
\thanks{This work was supported by the H2020 ERC Consolidator Grant L2C (Grant 864017), the CHIST-ERA 2018 project DRUID-NET, the Walloon Region and the Innoviris Foundation. R. J. is an FNRS Research Associate and a Fulbright Fellow.}
}

\author{Wei~Ren and Rapha\"el Jungers
\thanks{W. Ren and R. Jungers are with ICTEAM institute, UCLouvain, 1348 Louvain-la-Neuve, Belgium. Email: \texttt{\small w.ren@uclouvain.be, raphael.jungers@uclouvain.be}.}
}

\maketitle

\begin{abstract}
In this paper, we investigate the controller design problem for linear disturbed systems under signal temporal logic (STL) specifications imposing both spatial and temporal constraints on system behavior. We first implement zonotope-based techniques to partition the state space into finite cells, then propose an evaluation mechanism to rearrange the time constraints of the STL specification, and finally decompose the global STL formula into finite local STL formulas. In this way, each cell has a local control design problem, which is further formulated into a local optimization problem. To deal with each local optimization problem, we take advantage of the properties of zonotopes and reachability analysis to design local controller consisting of feedforward and feedback parts. By solving all local optimization problems, all local controllers are combined to guarantee the global STL specification. Finally, a numerical example is presented to illustrate the derived results.
\end{abstract}

\section{Introduction}
\label{sec-intro}

In modern control applications like autonomous driving and collaborative control \cite{Fong2001collaborative}, a recent trend is to consider complex task specifications instead of standard objectives including consensus, connectivity maintenance and collision avoidance. These complex task specifications motivate the need of an expressive language for specifying high-level objectives for dynamical systems \cite{Baier2008principles}. In this respect, temporal logics provide an intuitive way and a compact mathematical formalism to specify desired behaviors of dynamical systems for planning and control synthesis. For instance, linear temporal logic (LTL) has been widely used in formal verification \cite{Baier2008principles}. Toward temporal logic specifications, a general approach is to construct a symbolic abstraction for the considered system such that formal methods like automata-theoretic and graph-searching methods can be applied to design a discrete controller to ensure the satisfaction of LTL specifications \cite{Fainekos2009temporal, Belta2007symbolic, Pola2008approximately, Meyer2019hierarchical}. However, the abstraction-based control design may have huge computational complexity \cite{Reissig2017feedback}, which increases with system dimension and specification complexity, and is based on backward search techniques \cite{Tabuada2009verification}, which may be not available when time constraints are involved, e.g., in Signal Temporal Logic (STL).

STL is a real-time temporal logic defined over continuous signals \cite{Maler2004monitoring} and allows the specification of properties over dense-time. STL has the advantage of naturally admitting a quantitative semantics which, in addition to the boolean answer to the satisfaction question, provides a real number grading the quality of the satisfaction or violation \cite{Raman2014model}. This semantics is to assess the robustness of the systems to parameter or timing variations, which is different from LTL with a Boolean satisfaction only. Many approaches have been proposed to deal with control synthesis under STL specifications, such as control barrier function \cite{Garg2019control, Lindemann2020barrier}, optimization methods \cite{Raman2014model} and learning-based methods \cite{Aksaray2016q}. However, these approaches address global STL formulas \cite{Raman2014model} for the planning synthesis or assume the existence of local STL formulas \cite{Lindemann2020barrier} to facilitate the control synthesis. How to determine local STL formulas is unknown, and to the best of our knowledge, there are few works \cite{Charitidou2021signal} on how to decompose a global STL formula into local ones, which is still an open problem.

In this paper, we propose a novel framework for the optimal control synthesis of linear disturbed systems based on the decomposition of STL specifications. Toward this goal, the first step is to decompose the global STL formula into finite local ones. To be specific, we first partition the state space into finite zonotopes and constrained zonotopes, which are called cells and allowed to intersect with each other \cite{Ren2021zonotope}. Then, from the intersection relation among all cells, an undirected graph is derived to verify the LTL formula induced from the STL formula by taking the quantitative semantics as the Boolean semantics and ignoring time constraints \cite{Raman2014model}. Hence, the dissatisfaction of the STL-induced LTL formula implies the dissatisfaction of the STL formula, whereas not \emph{vice versa}, which provides a necessary condition for the satisfaction of the STL formula. Third, with the verification of the STL-induced LTL formula, we propose an evaluation mechanism to evaluate each cell and to decompose the time constraints in the STL formula. Finally, by combining the state space partition and the decomposition of time constraints, we yield the decomposition of the STL formula into finite local ones.

After decomposing the STL formula, a reachability-based synthesis approach is proposed to design the controllers for all local STL formulas, which is the second step. From the verification of the STL-induced LTL formula and the evaluation mechanism, we determine a sequence of cells with local STL formulas. In each cell, the intersection region between the current and previous cells is the initial region, the intersection region between the current and next cells is the final target region, and a local time constraint is arranged via the evaluation mechanism. This setting is further transformed into a local optimization problem involving both state and input constraints. In order to resolve each optimization problem, we first consider the nominal system to design the reference trajectory and the feedforward controller by taking advantage of the properties of (constrained) zonotopes, then address the disturbance case by designing the feedback controller to track the reference trajectory. In summary, comparing with \cite{Garg2019control} where the decomposition of STL formulas is only shown via a numerical example and \cite{Lindemann2020barrier} where local STL formulas are assumed to be given \emph{a priori}, the proposed approach provides a formal way to decompose the STL formula into finite local ones. In this way, intermediate goals are introduced formally comparing with \cite{Schurmann2020optimizing} where intermediate goals are used while how to introduce them is unknown. Finally, comparing with our previous work \cite{Ren2021zonotope} on LTL formulas, a further step is made in this paper towards the controller design for STL formulas.

The rest of this paper is organized below. The system and problem to be studied are introduced in Section \ref{sec-notepre}. The decomposition of STL specifications is proposed in Section \ref{sec-partition}. The reachability-based control strategy is presented in Section \ref{sec-optimalcontrol}. A numerical example is given in Section \ref{sec-example}. Conclusion and future work are presented in Section \ref{sec-conclusion}.

\section{Problem Formulation}
\label{sec-notepre}

Let $\mathbb{R}:=(-\infty, +\infty)$; $\mathbb{R}^{+}:=[0, +\infty)$; $\mathbb{N}:=\{0, 1, \ldots\}$; $\mathbb{N}^{+}:=\{1, 2, \ldots\}$. $\mathbb{R}^{n}$ denotes the $n$-dimensional Euclidean space. Given a vector $x\in\mathbb{R}^{n}$, $x_{i}$ is the $i$-th element of $x$, $\|x\|$ is the Euclidean norm of $x$, and $\|x\|_{\infty}$ is the infinity norm of $x$. Given two sets $\mathcal{A}$ and $\mathcal{B}$, $\mathcal{B}\backslash\mathcal{A}:=\{x: x\in\mathcal{B}, x\notin\mathcal{A}\}$. Given a compact set $\Lambda\subset\mathbb{R}^{n}$ and $\varepsilon>0$, $\Lambda^{\circ}$ is the interior of $\Lambda$; $\partial\Lambda$ is the boundary of $\Lambda$.

A set $Z\subset\mathbb{R}^{n}$ is a \emph{zonotope}, if there exists $(\mathbf{c}, \mathbf{G})\in\mathbb{R}^{n}\times\mathbb{R}^{n\times n_{g}}$ such that
\begin{equation}
\label{eqn-1}
Z:=\{\mathbf{c}+\mathbf{G}\xi: \|\xi\|_{\infty}\leq1\},
\end{equation}
where $\mathbf{c}\in\mathbb{R}^{n}$ is the center, and $\mathbf{G}\in\mathbb{R}^{n\times n_{g}}$ is the generator matrix with each column being a generator. The form \eqref{eqn-1} is called the generator-representation (G-representation) of the zonotope $Z$. A set $Y\subset\mathbb{R}^{n}$ is a \emph{constrained zonotope}, if there exists $(\mathbf{c}, \mathbf{G}, \mathbf{A}, \mathbf{b})\in\mathbb{R}^{n}\times\mathbb{R}^{n\times n_{g}}\times\mathbb{R}^{n_{c}\times n_{g}}\times\mathbb{R}^{n_{c}}$ such that
\begin{equation}
\label{eqn-2}
Y:=\{\mathbf{c}+\mathbf{G}\xi: \|\xi\|_{\infty}\leq1, \mathbf{A}\xi=\mathbf{b}\},
\end{equation}
where $\mathbf{A}\xi=\mathbf{b}$ is the constraint condition. The constrained zonotope is called constrained generator representation (CG-representation). We use the notation $Z=\{\mathbf{c}, \mathbf{G}\}$ for zonotopes, and $Y=\{\mathbf{c}, \mathbf{G}, \mathbf{A}, \mathbf{b}\}$ for constrained zonotopes. For a zonotope $Z=\{\mathbf{c}, \mathbf{G}\}$, if $\mathbf{G}$ is diagonal, orthogonal or invertible, then $Z$ is reduced to a box, a hypercube or a parallelotope, respectively. A convex polytope is a zonotope if and only if every 2-face is centrally symmetric \cite{Scott2016constrained}. From \cite{Scott2016constrained}, $Y\subset\mathbb{R}^{n}$ is a constrained zonotope if and only if it is a convex polytope. Given $Y=\{\mathbf{c}, \mathbf{G}, \mathbf{A}, \mathbf{b}\}$, $Y\neq\varnothing$ if and only if $\min\{\|\xi\|_{\infty}: \mathrm{A}\xi=\mathbf{b}\}\leq1$; $z\in Y$ if and only if $\min\{\|\xi\|_{\infty}: \mathbf{G}\xi=z-\mathbf{c}, \mathbf{A}\xi=\mathbf{b}\}\leq1$. These criteria are available for the zonotope \eqref{eqn-1} by removing $\mathbf{A}\xi=\mathbf{b}$.

\subsection{Signal Temporal Logic}
\label{subsec-approbisimu}

Signal Temporal Logic (STL) \cite{Maler2004monitoring} determines whether a predicate $\mu$ is true (i.e., $\top$) or false (i.e., $\perp$). The predicate $\mu$ is evaluated based on a continuously differentiable function $h: \mathbb{R}^{n}\rightarrow\mathbb{R}$ as follows: for $\boldsymbol{x}\in\mathbb{R}^{n}$,
$\mu:=\left\{\begin{aligned}
\top, &\text{ if } h(\boldsymbol{x})\geq0 \\
\perp, &\text{ if } h(\boldsymbol{x})<0.
\end{aligned}\right.$ The syntax of an STL formula $\phi$ is given by
\begin{align*}
\phi::=\top \mid \mu\mid \neg\phi \mid \phi_{1}\wedge\phi_{2} \mid \phi_{1}\textsf{U}_{[a, b]}\phi_{2},
\end{align*}
where $\phi, \phi_{1}, \phi_{2}$ are STL formulas and $a, b\in\mathbb{R}^{+}$ with $a\leq b$. Let $(\boldsymbol{x}, t)\models\phi$ denote the satisfaction of the formula $\phi$ by a signal $\boldsymbol{x}: \mathbb{R}^{+}\rightarrow\mathbb{R}^{n}$ at time $t$. The formula $\phi$ is satisfiable if $\exists\boldsymbol{x}: \mathbb{R}^{+}\rightarrow\mathbb{R}^{n}$ such that $(\boldsymbol{x}, t)\models\phi$. For a signal $\boldsymbol{x}: \mathbb{R}^{+}\rightarrow\mathbb{R}^{n}$, the STL semantics \cite{Maler2004monitoring} are recursively given below.
\begin{align*}
(\boldsymbol{x}, t)&\models\mu & \Leftrightarrow\ & h(\boldsymbol{x}(t)) \geq 0 \\
(\boldsymbol{x}, t)&\models\neg\phi & \Leftrightarrow\ & \neg((\boldsymbol{x}, t)\models\phi) \\
(\boldsymbol{x}, t)&\models\phi_{1}\wedge \phi_{2} & \Leftrightarrow\ & (\boldsymbol{x}, t)\models\phi_{1}\wedge(\boldsymbol{x}, t) \models \phi_{2} \\
(\boldsymbol{x}, t)&\models\phi_{1}\textsf{U}_{[a, b]} \phi_{2} & \Leftrightarrow\ & \exists t_{1} \in[t+a, t+b]\text { s.t. }(\boldsymbol{x}, t_{1})\models\phi_{2} \\
& && \wedge \forall t_{2}\in[t, t_{1}], (\boldsymbol{x}, t_{2}) \models \phi_{1} \\
(\boldsymbol{x}, t)&\models\textsf{F}_{[a, b]} \phi & \Leftrightarrow\ & \exists t_{1} \in[t+a, t+b] \text { s.t. }(\boldsymbol{x}, t_{1})\models\phi \\
(\boldsymbol{x}, t)&\models\textsf{G}_{[a, b]} \phi & \Leftrightarrow\ & \forall t_{1} \in[t+a, t+b],(\boldsymbol{x}, t_{1})\models\phi
\end{align*}
A signal $\boldsymbol{x}$ satisfies $\phi$, denoted by $\boldsymbol{x}\models\phi$ if $(\boldsymbol{x}, 0)\models\phi$.

\subsection{Problem Formulation}

In this paper, we focus on the following STL fragment
\begin{subequations}
\label{eqn-3}
\begin{align}
\label{eqn-3a}
\hat{\phi}&::=\textsf{G}_{[a, b]}\mu\mid\textsf{F}_{[a, b]}\mu, \\
\label{eqn-3b}
\phi&::=\wedge^{\boldsymbol{n}}_{i=1}\hat{\phi}_{i},
\end{align}
\end{subequations}
where $0\leq a\leq b$, $\boldsymbol{n}\in\mathbb{N}$ is finite, and $\hat{\phi}_{i}$ is of the form \eqref{eqn-3a}. The STL fragment \eqref{eqn-3} is expressive enough to include the operator $\textsf{U}_{[a, b]}$. In particular, $\mu_{1}\textsf{U}_{[a, b]}\mu_{2}$ can be written equally as $\textsf{G}_{[a, t']}\mu_{1}\wedge\textsf{F}_{[t', t']}\mu_{2}$ for certain $t'\in[a, b]$; see \cite{Charitidou2021signal}.

Consider the following linear control system
\begin{equation}
\label{eqn-4}
\dot{x}(t)=Ax(t)+Bu(t)+Cw(t),
\end{equation}
where $x(t)\in\mathbb{X}\subset\mathbb{R}^{n}$ is the system state, $u(t)\in\mathbb{U}\subset\mathbb{R}^{m}$ is the control input, $w(t)\in\mathbb{W}\subset\mathbb{R}^{p}$ is the external disturbance, and $A, B, C$ are matrices with appropriate dimensions. Assume that the sets $\mathbb{X}, \mathbb{U}, \mathbb{W}$ are convex and compact. For the system \eqref{eqn-4}, its solution at the time $t\in\mathbb{R^{+}}$ is denoted as $\mathbf{x}(t, x_{0}, u, w)$ with the initial state $x_{0}\in\mathbb{X}$, the control input $u\in\mathbb{U}$ and disturbance $w\in\mathbb{W}$. Given an initial set $\mathbb{X}_{0}\subset\mathbb{X}$ and the time $t\in\mathbb{R^{+}}$, the reachable set starting from an initial set $\mathbb{X}_{0}\subset\mathbb{X}$ is defined as $\mathcal{R}(t, \mathbb{X}_{0}):=\{x(t)\in\mathbb{X}: \exists x_{0}\in\mathbb{X}_{0}, u\in\mathbb{U}, w\in\mathbb{W} \text{ such that } \mathbf{x}(t, x_{0}, u, w)=x(t)\}$. Hence, $\mathbf{x}(0, x_{0}, u, w)=x_{0}$ and $\mathcal{R}(0, \mathbb{X}_{0})=\mathbb{X}_{0}$. In this paper, the problem to be studied is formulated below.

\begin{problem}
\label{prob-1}
Consider the system \eqref{eqn-4} and a desired specification expressed as an STL formula $\phi$ of the form \eqref{eqn-3}. Given an initial set $\mathbb{X}_{0}\subset\mathbb{X}$, find a control design strategy such that the STL formula $\phi$ can be satisfied for the system \eqref{eqn-4}.
\end{problem}

To solve Problem \ref{prob-1}, we propose a two-step control strategy. In the first step, we partition the state space $\mathbb{X}\subset\mathbb{R}^{n}$ and decompose the time constraints (i.e., $[a, b]$ in \eqref{eqn-3a}) in the STL formula $\phi$. Hence, the global STL formula can be decomposed into finite local ones in Section \ref{sec-partition}. With the formula decomposition, the second step is to develop an iterative algorithm to design local controllers for all local STL formulas in Section \ref{sec-optimalcontrol}. If all local controllers exist, then they can be combined together such that $\phi$ is satisfied for the system \eqref{eqn-4}.

\section{Decomposition of STL Specifications}
\label{sec-partition}

This section is devoted to the first step of the propose control strategy. To be specific, to decompose the STL specification into finite local ones, we first partition the state space via zonotope techniques in Section \ref{subsec-strategy}, then propose an evaluation-function-based approach to decompose the time constraints in the STL specification in Section \ref{subsec-decomtime}, and finally derive the decomposition of the STL specification in Section \ref{subsec-localSTL}.

\subsection{Partition of State Space}
\label{subsec-strategy}

The partition strategy is presented in Algorithm \ref{alg-1} and has two steps. The first step is to generate finite zonotopes and constrained zonotopes to cover the state space $\mathbb{X}\subset\mathbb{R}^{n}$ (i.e., line 1 in Algorithm \ref{alg-1}), and the generation rules will be presented in detail in the next subsection.

\begin{algorithm}[!t]
\DontPrintSemicolon
\small
\caption{State Space Partition}
\label{alg-1}
\KwIn{the state space $\mathbb{X}\subset\mathbb{R}^{n}$, the parameter $\varepsilon>0$}
\KwOut{the partition $\mathbf{Z}$ of the state space $\mathbb{X}$}
Generate finite zonotopes $Z_{i}$ and constrained zonotopes $Y_{j}$ to cover the state domain $\mathbb{X}$\;
Expand both $Z_{i}$ and $Y_{j}$ via the parameter $\varepsilon>0$\;
\textbf{return} $\mathbf{Z}=(\cup^{N}_{i=1}\mathbf{E}_{\varepsilon}(Z_{i}))\cup(\cup^{M}_{j=1}\mathbf{E}_{\varepsilon}(Y_{j}))$
\end{algorithm}

\subsubsection{Generation of Zonotopes and Constrained Zonotopes}

How to generate zonotopes and constrained zonotopes is presented in Algorithm \ref{alg-2}. The generation rule consists of two parts: the first part is to generate zonotopes (lines 1-3), and the second part is to generate constrained zonotopes (lines 4-11) based on the zonotope construction.

To begin with, the number of zonotopes to be generated is set \emph{a priori} as $N\in\mathbb{N}$, and we choose $N$ points $\mathbf{c}_{i}\in\mathbb{X}$ arbitrarily as the centers of zonotopes to be generated, where $i\in\mathcal{N}_{1}:=\{1, \ldots, N\}$. Here, we assume that $N\in\mathbb{N}$ satisfies $N>n$ with $n\in\mathbb{N}$ being the dimension of the state space. Second, we connect these centers such that each center is connected with at least $n\in\mathbb{N}$ neighbouring centers. Therefore, for each center, these connections lead to at least $n\in\mathbb{N}$ vectors, which will be further used as the generators for each zonotope. Finally, with these centers and generators, we can generate $N$ zonotopes as in line 3 in Algorithm \ref{alg-2}.

In Algorithm \ref{alg-2}, the constraints on the number $N\in\mathbb{N}$ and the matrix $\mathbf{G}_{i}$ are to guarantee the well-constructedness of zonotopes; see \cite{Ren2021zonotope} for more details. In particular, the connection of each center with at least $n$ neighbouring centers leads to the full-rank matrix $\mathbf{G}_{i}$, which also implies $N>n$. If $\mathbf{G}_{i}$ is not full-rank, then the dimension of the generated zonotope is less than $n$, which shows that the zonotope is not well-constructed. On the other hand, different values of $N, k_{i}\in\mathbb{N}$ in \eqref{eqn-5} result in different numbers of zonotopes and generators. Especially, different $N$ and different generators result in different zonotopes.

\begin{algorithm}[!t]
\DontPrintSemicolon
\small
\caption{Zonotope Generation}
\label{alg-2}
\KwIn{the state space $\mathbb{X}\subset\mathbb{R}^{n}$, the integer $N>n$}
\KwOut{finite zonotopes and constrained zonotopes}
Choose $\{\mathbf{c}_{i}\in\mathbb{X}: i=1, \ldots, N\}$ randomly\;
{Connect these points to generate full-rank matrices
\begin{equation}
\label{eqn-5}
\mathbf{G}_{i}=0.5(\mathbf{c}_{k_{1}}-\mathbf{c}_{i}, \ldots, \mathbf{c}_{k_{i}}-\mathbf{c}_{i}), \quad k_{i}\geq n
\end{equation}}\;
\vspace*{-\baselineskip}
{Construct the zonotope $Z_{i}=\{\mathbf{c}_{i}+\mathbf{G}_{i}\xi: \|\xi\|_{\infty}\leq1\}$}\;
\eIf{$\mathbb{X}\setminus(\cup^{N}_{i=1}Z_{i})=\varnothing$}{
No need to construct constrained zonotopes
}{
Generate the set $\mathbb{V}_{1}$ from the intersection among $\partial Z_{1}, \ldots, \partial Z_{N}, \partial\mathbb{X}$\;
Refine $\mathbb{V}_{1}$ as $\mathbb{V}$ by excluding those in $\cup^{N}_{i=1}Z^{\circ}_{i}$\;
Generate the region set $\mathbb{S}_{1}$ from all vertices in $\mathbb{V}$\;
Refine $\mathbb{S}_{1}$ as $\mathbb{S}$ by excluding those in $\cup^{N}_{i=1}Z_{i}$\;
Construct $M$ constrained zonotopes $\cup^{M}_{j=1}Y_{j}\supseteq\mathbb{S}$\;
}
\textbf{return} $(\cup^{N}_{i=1}Z_{i})\cup(\cup^{M}_{j=1}Y_{j})$
\end{algorithm}

If the union of all generated zonotopes covers the whole state space, then there is not need for the generation of constrained zonotopes; otherwise, the generation of constrained zonotopes is needed. Note that constrained zonotopes are asymmetric \cite{Scott2016constrained, Rego2020guaranteed} and can be used to cover asymmetric regions. The construction rule is based on the intersection points among the generated zonotopes and the state space, and presented explicitly as follows. First, from the definition of zonotopes, we derive the vertices of each generated zonotope, and these vertices are connected to form the boundaries (i.e., $\partial Z_{i}$ with $i\in\mathcal{N}_{1}$) of each generated zonotope. Since each generated zonotope may intersect with its neighbour zonotopes or the boundaries of the state space, we can check the relation of the boundaries of all generated zonotopes and the state space to determine all intersection points (i.e. the set $\mathbb{V}_{1}$ in line 7). Second, since these intersection points may be contained in some generated zonotopes, we exclude those in the interiors of at least one of all generated zonotopes, which refines $\mathbb{V}_{1}$ as $\mathbb{V}$ in line 8. Third, all elements in $\mathbb{V}$ are used to generate all regions forming the set $\mathbb{S}_{1}$ in line 9, and similar to line 8, the set $\mathbb{S}_{1}$ is refined by excluding the regions in generated zonotopes in line 10. Finally, these regions are transformed into the form of constrained zonotopes (line 11).

The generation of constrained zonotopes depends on basic operations of zonotopes. To be specific, the vertices of all generated zonotopes can be obtained via the transformation of zonotopes from the G-representation into the V-representation \cite[Algorithm 2]{Kochdumper2019representation}. These vertices are connected to determine the intersection vertices to form the set $\mathbb{V}_{1}$ in line 7 of Algorithm \ref{alg-2}. The set $\mathbb{V}_{1}$ is refined as $\mathbb{V}$ in line 8 by excluding those in the generated zonotopes. Based on the vertices in $\mathbb{V}$ and using the transformation from V-representation into Z-representation in \cite{Kochdumper2019representation}, constrained zonotopes are generated in lines 9-11 of Algorithm \ref{alg-2} to cover the region $\mathbb{X}\setminus(\cup^{N}_{i=1}Z_{i})$.

\subsubsection{Expansion of Zonotopes and Constrained Zonotopes}

Since the generated zonotopes and constrained zonotopes are not necessarily overlapped, the second step of Algorithm \ref{alg-1} is to implement the expansion operator to expand all generated zonotopes and constrained zonotopes. To be specific, given a constant $\varepsilon>0$, the $\varepsilon$-expansions of $Z=\{\mathbf{c}, \mathbf{G}\}$ and $Y=\{\mathbf{c}, \mathbf{G}, \mathbf{A}, \mathbf{b}\}$ are defined as $\mathbf{E}_{\varepsilon}(Z):=\{\mathbf{c}, (1+\varepsilon)\mathbf{G}\}$ and $\mathbf{E}_{\varepsilon}(Y):=\{\mathbf{c}, (1+\varepsilon)\mathbf{G}, \mathbf{A}, (1+\varepsilon)\mathbf{b}\}$. It is easy to check that $\mathbf{E}_{\varepsilon}(Z)$ and $\mathbf{E}_{\varepsilon}(Y)$ are still zonotope and constrained zonotope, respectively. The expansion operation ensures each (constrained) zonotope to overlap with its neighbouring zonotopes and constrained zonotopes. Therefore, Algorithm \ref{alg-1} produces the union of finite zonotopes and constrained zonotopes to partition the state space. Different from the existing partition methods in \cite{Belta2007symbolic, Fainekos2009temporal}, the proposed method allows each cell (i.e., zonotope or constrained zonotope) to intersect with its neighbours. We emphasize that such intersection relation will play an important role in the controller synthesis afterwards.

All generated zonotopes and constrained zonotopes are labeled via a set of symbols $\Pi:=\{\pi_{1}, \ldots,$ $\pi_{N}, \pi_{N+1}, \ldots,$ $\pi_{N+M}\}$, where $\llbracket\pi_{i}\rrbracket:=\mathbf{E}_{\varepsilon}(Z_{i})$ for $i\in\mathcal{N}_{1}$ and $\llbracket\pi_{N+j}\rrbracket:=\mathbf{E}_{\varepsilon}(Y_{j})$ for $j\in\mathcal{N}_{2}:=\{1, \ldots, M\}$. Let $\mathcal{N}:=\{1, \ldots, N+M\}$ and the state space is partitioned as
\begin{align}
\label{eqn-6}
\mathbf{Z}&:=\{\mathbf{Z}_{k}: \mathbf{Z}_{i}=\llbracket\pi_{k}\rrbracket, k\in\mathcal{N}\}.
\end{align}

\begin{algorithm}[!t]
\DontPrintSemicolon \small
\caption{Adjacency Matrix}
\label{alg-3}
\KwIn{the partition $\mathbf{Z}$, the forbidden region $\mathcal{O}\subset\mathbb{X}$}
\KwOut{the matrix $\Upsilon=[a_{ij}]$ and the set $\mathbf{I}=\{\mathbf{I}_{ij}\}$}
\For{$i=1:1:N+M-1$}{
\eIf{$\mathbf{Z}_{i}\cap\mathcal{O}\neq\mathbf{Z}_{i}$}{
\For{$j=1:1:N+M$}{
\eIf{$\mathbf{Z}_{j}\cap\mathcal{O}\neq\mathbf{Z}_{j}$}{
$\Omega=\mathbf{Z}_{i}\cap\mathbf{Z}_{j}$\;
\eIf{$\Omega=\varnothing$}{
$a_{ij}=0$\;}{
\eIf{$\Omega\setminus\mathcal{O}$ is a connected region}{
$a_{ij}=1$ and $\mathbf{I}_{ij}=\Omega$\;
}{$a_{ij}=0$\;}
}
}{$a_{ij}=0$\;}
}
}{$a_{ij}=0$ for $j\in\mathcal{N}$\;}
}
\textbf{return} $\Upsilon=[a_{ij}]$ and $\mathbf{I}=\{\mathbf{I}_{ij}\}$
\end{algorithm}

\subsubsection{Topological Graph}
\begin{figure}[!t]
\begin{center}
\begin{picture}(75, 75)
\put(-70, -12){\resizebox{75mm}{30mm}{\includegraphics[width=2.5in]{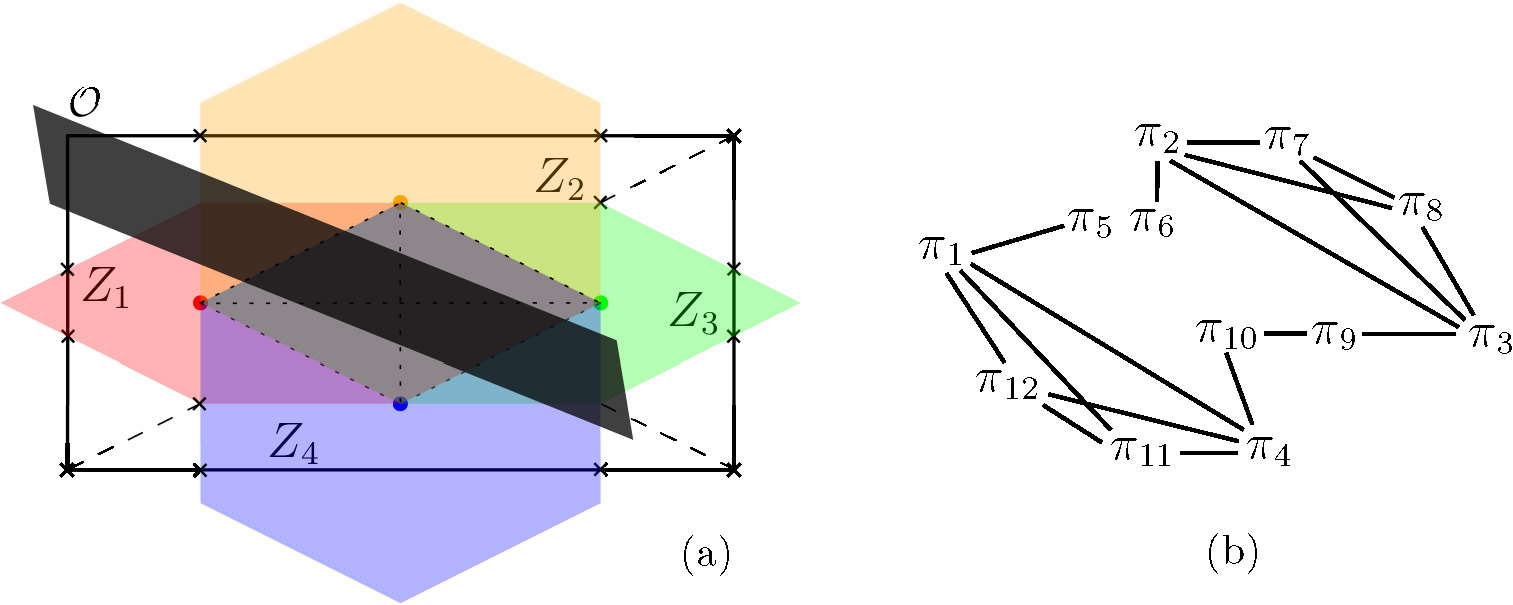}}}
\end{picture}
\end{center}
\caption{Illustration of Algorithm \ref{alg-3}. (a) The partition of the state space $\mathbb{X}\subset\mathbb{R}^{2}$, where the black region is the obstacle $\mathcal{O}\subset\mathbb{X}$. (b) The generated graph $\mathcal{G}=(\mathcal{V}, \mathcal{E})$ with the symbol set $\Pi:=\{\pi_{1}, \ldots, \pi_{12}\}$.}
\label{fig-1}
\end{figure}

Based on the intersection relation among all cells (i.e., zonotopes and constrained zonotopes), we can obtain a undirected graph. First, some auxiliary notations are introduced. Given the state space $\mathbb{X}\subset\mathbb{R}^{n}$, all forbidden states including obstacles and states that are not allowed to be visited consist of the set $\mathcal{O}:=\{\mathcal{O}_{k}: k\in\mathbb{K}\}\subset\mathbb{X}$, where $\mathbb{K}\subseteq\mathbb{N}$ is a finite index set. A region $A\subset\mathbb{X}$ is \emph{admissible}, if $A\setminus\mathcal{O}$ is a nonempty and connected region. To derive the graph, the first step is to show the existence of an edge between any two cells by verifying if their intersection is admissible. After verifying the intersection relation, we have all admissible intersection regions and the corresponding adjacency matrix, which are presented in Algorithm \ref{alg-3}. From finite number of all cells and the analysis in \cite{Kochdumper2020sparse}, the termination time of Algorithm \ref{alg-3} is finite. The derived matrix $\Upsilon$ is transformed into an undirected graph $\mathcal{G}=(\mathcal{V}, \mathcal{E})$ with $\mathcal{V}=\Pi$ and the edge set $\mathcal{E}\subseteq\Pi\times\Pi$, where $(\pi_{i}, \pi_{j})\in\mathcal{E}$ if $a_{ij}=1$; see Fig. \ref{fig-1}.

\subsubsection{Admissible path for the STL-induced LTL}

By treating predicates as Boolean variables and ignoring time constraints, the STL formula $\phi$ can be reduced to the classic LTL formula \cite{Raman2014model}. The STL-induced LTL formula is denoted as $\phi_{\textsf{LTL}}$. From $\phi_{\textsf{LTL}}$, the set of regions of interest is derived and denoted as the set of finite constrained zonotopes $\mathbf{Z}_{\textsf{LTL}}=\{Y_{\mathrm{r}j}: j\in\mathbb{J}\}$ with the symbol set $\{\pi_{\mathrm{r}1}, \ldots, \pi_{\mathrm{rJ}}\}$, where $\mathbb{J}:=\{1, \ldots, \mathrm{J}\}$ is the index set and $\mathrm{J}\in\mathbb{N}$. The initial state region is assumed to be contained in a constrained zonotope $Y_{0}$ with the symbol $\pi_{0}$. Both $Y_{0}$ and $Y_{\mathrm{r}j}$ may intersect with some elements of $\mathbf{Z}$ in \eqref{eqn-6}, which can be verified easily. Therefore, the graph $\mathcal{G}$ is generalized as $\bar{\mathcal{G}}=(\bar{\mathcal{V}}, \bar{\mathcal{E}})$, where the vertex set is $\bar{\mathcal{V}}=\Pi\cup\{\pi_{0}, \pi_{\mathrm{r}1}, \ldots, \pi_{\mathrm{r}\mathrm{J}}\}$, and the edge set is $\bar{\mathcal{E}}\subseteq\bar{\mathcal{V}}\times\bar{\mathcal{V}}$ extending $\mathcal{E}$ by adding the intersection relations among $Y_{0}, Y_{\mathrm{r}j}$ and $\mathbf{Z}$. For the LTL formula $\phi_{\textsf{LTL}}$, any standard LTL model checker \cite{Baier2008principles} can be applied to solve the LTL planning problem. Hence, we can obtain an accepting path $\bar{\pi}:=\bar{\pi}_{0}\bar{\pi}_{1}\bar{\pi}_{2}\ldots$ with $\bar{\pi}_{0}=\pi_{0}$ and $\bar{\pi}_{k}\in\{\pi_{\mathrm{r}1}, \ldots, \pi_{\mathrm{r}\mathrm{J}}\}$, $k\in\mathbb{N}^{+}$.

In the graph $\bar{\mathcal{G}}$, each path $\mathbf{P}$ can be projected into a sequence of cells $\mathbf{Z}_{\textsf{p}}:=\{\mathbf{Z}(\pi):=\llbracket\pi\rrbracket\in\mathbf{Z}: \pi\in\mathbf{P}\}$. The intersection region set in $\mathbf{Z}_{\textsf{p}}$ is denoted as $\mathbf{I}_{\textsf{p}}:=\{\mathbf{I}^{\textsf{p}}_{ij}=\mathbf{Z}_{\textsf{p}i}\cap\mathbf{Z}_{\textsf{p}j}\in\mathbf{I}: \forall \mathbf{Z}_{\textsf{p}i}, \mathbf{Z}_{\textsf{p}j}\in\mathbf{Z}_{\textsf{p}}\}$. With the graph $\bar{\mathcal{G}}$, the following theorem is derived to verify the satisfaction of the LTL formula $\phi_{\textsf{LTL}}$.

\begin{theorem}
\label{thm-1}
Consider the state space $\mathbb{X}\subseteq\mathbb{R}^{n}$, the initial state region $\mathbb{X}_{0}\subset Y_{0}$, the obstacle $\mathcal{O}\subset\mathbb{X}$ and the LTL formula $\phi_{\textsf{LTL}}$. The following two statements are equivalent.
\begin{enumerate}
  \item The formula $\phi_{\textsf{LTL}}$ can be satisfied in $\mathbb{X}\subseteq\mathbb{R}^{n}$.
  \item In the graph $\bar{\mathcal{G}}$ there exists a path $\mathbf{P}$ realizing the accepting path $\bar{\pi}:=\bar{\pi}_{0}\bar{\pi}_{1}\bar{\pi}_{2}\ldots$. That is, there exists a sub-path in $\mathbf{P}$ for any pair $(\bar{\pi}_{i}, \bar{\pi}_{i+1})$ with $i\in\mathbb{N}$.
\end{enumerate}
\end{theorem}

If the path $\mathbf{P}$ exists, it is not necessarily unique. Theorem \ref{thm-1} shows the satisfiability of the LTL formula $\phi_{\textsf{LTL}}$ since $\phi_{\textsf{LTL}}$ only depends on the state space. Due to the time constraints, the satisfactability of the STL formula $\phi$ involves the system dynamics. Therefore, the satisfiability of the LTL formula $\phi_{\textsf{LTL}}$ does not imply the satisfiability of the STL formula $\phi$, while the STL formula $\phi$ cannot be satisfied if the LTL formula $\phi_{\textsf{LTL}}$ is not satisfiable. That is, Theorem \ref{thm-1} only provides necessary conditions for the satisfiability of the STL formula $\phi$.

\subsection{Decomposition of Time Constraints}
\label{subsec-decomtime}

To derive local STL formulas, the second step is to decompose the time constraints in the STL formula $\phi$ using the state space partition. Such decomposition depends on the evaluation of each cell, which is defined as a function $\evl: 2^{\mathbb{X}}\rightarrow\mathbb{R}^{+}$ satisfying $\evl(v)=0$ if and only if $v=\varnothing$. A simple and direct evaluation of each cell is based on its volume. Specifically, for each $\mathbf{Z}_{i}\in\mathbf{Z}$, we define by $\vol(\mathbf{Z}_{i})$ the volume of $\mathbf{Z}_{i}$, and further $\evl(\mathbf{Z}_{i})=\vol(\mathbf{Z}_{i})$. If $\mathbf{Z}_{i}$ is a zonotope, then from \cite[Section 3]{Gover2010determinants} we have $\vol(\mathbf{Z}_{i})=\sqrt{\det(\mathbf{G}_{i}\mathbf{G}^{\top}_{i})}$, where $\mathbf{G}_{i}$ is the generator matrix and $\det(\cdot)$ is the determinant operator. The volume of constrained zonotopes can be computed via different methods; see e.g., \cite{Lawrence1991polytope}. On the other hand, other indices can be included to evaluate each cell, such as the constraints in $\phi$ and the volume of the admissible region in each cell, which deserves further study.

From Theorem \ref{thm-1}, we obtain finite admissible paths for the STL-induced LTL formula $\phi_{\textsf{LTL}}$. For each path $\mathbf{P}_{i}$, we derive a sequence of finite cells denoted as $\{\mathbf{Z}_{ij}: \mathbf{Z}_{ij}\in\mathbf{Z}, j\in\mathbb{L}_{i}\}$ with the index set $\mathbb{L}_{i}\subset\mathbb{N}$, and have the evaluation of this path, i.e., $\evl(\mathbf{P}_{i})=\sum_{j\in\mathbb{L}_{i}}\evl(\mathbf{Z}_{ij})$. By comparing all these evaluations, we choose the optimal path (denoted by $\mathbf{P}^{\ast}$) with the minimal value $\evl(\mathbf{P}^{\ast}_{i})$. That is, the path $\mathbf{P}^{\ast}$ is established via the following optimization problem.
\begin{subequations}
\label{eqn-7}
\begin{align}
\label{eqn-7a}
\min& \  \sum_{j\in\mathbb{L}_{i}}\evl(\mathbf{Z}_{ij}) \\
\label{eqn-7b}
\text{s.t.}
&\quad   \{\mathbf{Z}_{ij}: j\in\mathbb{L}_{i}\} \text{ is projected from } \mathbf{P}_{i}, \\
\label{eqn-7c}
&\quad \mathbf{P}_{i} \text{ is derived from Theorem \ref{thm-1}}.
\end{align}
\end{subequations}
From \eqref{eqn-7}, we expect the time constraints of the STL formula $\phi$ to be satisfied via the path $\mathbf{P}^{\ast}$.

With the optimal path $\mathbf{P}^{\ast}$, we have the set $\mathbf{Z}^{\ast}:=\{\mathbf{Z}^{\ast}_{i}: \mathbf{Z}^{\ast}_{i}\in\mathbf{Z}, i\in\mathbb{L}^{\ast}\}$ and the evaluation $\evl(\mathbf{P}^{\ast})$. Given an STL formula $\hat{\phi}$ as in \eqref{eqn-3a} with the time constraint $[a, b]\subset\mathbb{R}^{+}$ and the corresponding set $\mathbf{Z}^{\ast}$, we rearrange local time constraints as follows. Define
\begin{equation*}
\mathfrak{t}_{i}:=\frac{\evl(\mathbf{Z}^{\ast}_{i})}{\sum_{i\in\mathbb{L}^{\ast}}\evl(\mathbf{Z}^{\ast}_{i})}\mathfrak{h}(a, b),
\end{equation*}
where the function $\mathfrak{h}: \mathbb{R}^{+}\times\mathbb{R}^{+}\rightarrow(0, b]$ depends on the temporal operator. Hence, the time constraint for the cell $\mathbf{Z}^{\ast}_{i}$ is $\mathbb{T}_{i}=[\mathfrak{t}_{i-1}, \mathfrak{t}_{i-1}+\mathfrak{t}_{i}]$ with $\mathfrak{t}_{0}=0$ and $i\in\mathbb{L}^{\ast}$.

\subsection{From Global STL Formula to Local STL Formulas}
\label{subsec-localSTL}

With the optimal path $\mathbf{P}^{\ast}$ and the decomposition of time constraints, we decompose the global STL formula $\phi$ into finite local ones in this subsection. Let the set $\mathbf{Z}^{\ast}\subset\mathbf{Z}$ be from $\mathbf{P}^{\ast}$. For each pair $(\bar{\pi}_{i}, \bar{\pi}_{i+1})$ from the accepting path $\bar{\pi}$, there exists a sub-path $\mathbf{P}^{\ast}(\bar{\pi}_{i}, \bar{\pi}_{i+1})$ in $\mathbf{P}^{\ast}$ to connect $\bar{\pi}_{i}$ and $\bar{\pi}_{i+1}$, and this sub-path results in a subset $\mathbf{Z}^{\ast}(\bar{\pi}_{i}, \bar{\pi}_{i+1}):=\{\bar{\mathbf{Z}}^{\ast}_{1}, \ldots, \bar{\mathbf{Z}}^{\ast}_{\imath}\}$ of $\mathbf{Z}^{\ast}$ with finite $\imath\in\mathbb{N}$. The system \eqref{eqn-4} is expected to stay in $\mathbf{Z}^{\ast}(\bar{\pi}_{i}, \bar{\pi}_{i+1})$ and to move from $\bar{\mathbf{Z}}^{\ast}_{1}\cap\mathbf{Z}^{\ast}(\bar{\pi}_{i})$ to $\bar{\mathbf{Z}}^{\ast}_{\imath}\cap\mathbf{Z}^{\ast}(\bar{\pi}_{i+1})$. In this respect, in each cell $\mathbf{Z}^{\ast}_{i}\in\mathbf{Z}^{\ast}$ there exists a local LTL formula (denoted by $\bar{\varphi}_{i}$) consisting of three parts: a) the system state is in the state space $\mathbb{X}$ while avoiding the obstacle set $\mathcal{O}$; b) the system reaches the local target region $\mathbf{T}^{\ast}_{i}:=\mathbf{Z}^{\ast}_{i}\cap\mathbf{Z}^{\ast}_{i+1}$ or $\mathbf{T}^{\ast}_{i}:=\mathbf{Z}^{\ast}_{i}\cap\mathbf{Z}^{\ast}_{i+1}\cap\mathbf{Z}^{\ast}(\bar{\pi})$ if nonempty; c) if $\mathbf{Z}^{\ast}_{i}\cap\mathbf{Z}^{\ast}(\bar{\pi})\neq\varnothing$ and $\mathbf{T}^{\ast}_{i}\cap\mathbf{Z}^{\ast}(\bar{\pi})=\varnothing$, then the system visits the region $\mathbf{Z}^{\ast}_{i}\cap\mathbf{Z}^{\ast}(\bar{\pi})$. These three parts are denoted as the LTL formula $\bar{\varphi}_{i1}, \bar{\varphi}_{i2}, \bar{\varphi}_{i3}$, respectively. Note that $\bar{\varphi}_{i3}$ is neither unique nor always existent.

With these three parts, the local STL formula for $\mathbb{Z}^{\ast}_{i}$ can be defined as the form of \eqref{eqn-3}. To be specific, the three cases in \eqref{eqn-3} are addressed. For an STL formula $\phi$, let $\bar{\pi}$ be the accepting path for $\phi_{\textsf{LTL}}$ and $\mathbf{Z}^{\ast}(\bar{\pi}):=\{\mathbf{Z}^{\ast}_{1}, \ldots, \mathbf{Z}^{\ast}_{\imath}\}$ with $\imath\in\mathbb{N}$.
\begin{myenumi}
\item[\emph{Case 1:}] $\phi=\textsf{G}_{[a, b]}\varphi$. Let $\mathbf{R}(\varphi)$ be the region of interest from the formula $\varphi$ (see \cite{Meyer2019hierarchical, Garg2019control}), and thus $\mathbf{Z}^{\ast}_{\imath}\cap\mathbf{R}(\varphi)\neq\varnothing$. The time interval $[0, a]$ is rearranged to $\{\mathbf{Z}^{\ast}_{1}, \ldots, \mathbf{Z}^{\ast}_{\imath-1}\}$ such that $\mathbf{Z}^{\ast}_{\imath-1}\cap\mathbf{Z}^{\ast}_{\imath}$ can be reached at the time not later than $a$. Hence, $\bar{\phi}_{i}=\textsf{G}_{\mathbb{T}_{i}}\bar{\varphi}_{i1}\wedge\textsf{F}_{\mathbb{T}_{i}}\bar{\varphi}_{i2}$ for $i\in\{1, \ldots, \imath-1\}$ and       $\bar{\phi}_{\imath}:=\textsf{G}_{\mathbb{T}_{\imath}}\bar{\varphi}_{\imath1}\wedge\textsf{G}_{[a, b]}\bar{\varphi}_{\imath3}$ with $[a, b]\subseteq\mathbb{T}_{\imath}$.

  \item[\emph{Case 2:}] $\phi=\textsf{F}_{[a, b]}\varphi$. Similar to Case 1, $\mathbf{Z}^{\ast}_{\imath}\cap\mathbf{R}(\varphi)\neq\varnothing$, and $[0, b]$ is rearranged to $\{\mathbf{Z}^{\ast}_{1}, \ldots, \mathbf{Z}^{\ast}_{\imath-1}\}$ such that $\mathbf{Z}^{\ast}_{\imath-1}\cap\mathbf{Z}^{\ast}_{\imath}$ is reached at a time in $[a, b]$. Hence, $\bar{\phi}_{i}=\textsf{G}_{\mathbb{T}_{i}}\bar{\varphi}_{i1}\wedge\textsf{F}_{\mathbb{T}_{i}}\bar{\varphi}_{i2}$ for $i\in\{1, \ldots, \imath-1\}$ and       $\bar{\phi}_{\imath}:=\textsf{G}_{\mathbb{T}_{\imath}}\bar{\varphi}_{\imath1}\wedge\textsf{F}_{[a, b]}\bar{\varphi}_{\imath3}$ with $[a, b]\cap\mathbb{T}_{\imath}\neq\varnothing$.

  \item[\emph{Case 3:}] \emph{the conjunction of temporal operators}. The local formula is a direct extension of Cases 1 and 2. For instance, consider $\phi=\textsf{G}_{[a_{1}, b_{1}]}\varphi_{1}\wedge\textsf{F}_{[a_{2}, b_{2}]}\varphi_{2}$ with $\mathbf{Z}^{\ast}_{i_{1}}\cap\mathbf{R}(\varphi_{1})\neq\varnothing$ and $\mathbf{Z}^{\ast}_{i_{2}}\cap\mathbf{R}(\varphi_{2})\neq\varnothing$. If $\mathbf{Z}^{\ast}_{i_{1}}\neq\mathbf{Z}^{\ast}_{i_{2}}$ with $i_{1}<i_{2}=\imath$, then we follow Case 1 to construct $\bar{\phi}_{j}$ for $\textsf{G}_{[a_{1}, b_{1}]}\varphi_{1}$, where $j\in\{1, \ldots, i_{1}\}$. In particular, $\bar{\phi}_{i_{1}}:=\textsf{G}_{\mathbb{T}_{i_{1}}}\bar{\varphi}_{i_{1}1}
      \wedge\textsf{F}_{\mathbb{T}_{i_{1}}}\bar{\varphi}_{i_{1}2}\wedge\textsf{G}_{[a_{1}, b_{1}]}\bar{\varphi}_{i_{1}3}$ with $[a_{1}, b_{1}]\subseteq\mathbb{T}_{i_{1}}$. With $\mathbf{Z}^{\ast}_{i_{1}}$ and $\mathbb{T}_{i_{1}}$ as the initial conditions, we follow Case 2 to construct $\bar{\phi}_{j}$ for $\textsf{F}_{[a_{2}, b_{2}]}\varphi_{2}$ with $j\in\{i_{1}+1, \ldots, \imath\}$. If $\mathbf{Z}^{\ast}_{i_{1}}=\mathbf{Z}^{\ast}_{i_{2}}$, then we compare $[a_{1}, b_{1}]$ and $[a_{2}, b_{2}]$ to construct $\bar{\phi}_{i}$ as in Cases 1 and 2. The last local formula is $\bar{\phi}_{\imath}:=\textsf{G}_{\mathbb{T}_{\imath}}\bar{\varphi}_{\imath1}\wedge\textsf{G}_{[a_{1}, b_{1}]}\bar{\varphi}'_{\imath3}\wedge\textsf{F}_{[a_{2}, b_{2}]}\bar{\varphi}''_{\imath3}$ with $[a_{1}, b_{1}]\subseteq\mathbb{T}_{\imath}$ and $[a_{2}, b_{2}]\cap\mathbb{T}_{\imath}\neq\varnothing$. Similar construction can be finished for the case $i_{2}<i_{1}=\imath$. The available freedom in constructing $\bar{\phi}_{i}$ from the above conditions results in a non-unique constructive procedure.
\end{myenumi}
As a result, we decompose the global STL formula in \eqref{eqn-3} into finite local STL formulas.

\section{Reachability-based Controller Design}
\label{sec-optimalcontrol}

With the decomposition of the STL formula, we next design the controller such that the STL formula is satisfied by the system \eqref{eqn-4}. The proposed controller is given below
\begin{align}
\label{eqn-9}
u(t)=u_{\textmd{f}}(t, x_{0})+u_{\textmd{b}}(t, x_{\textmd{f}}(t)), \quad  x_{0}\in\mathbb{X}_{0}, t>0,
\end{align}
where $u_{\textmd{f}}$ is the feedforward controller to generate the reference trajectory $x_{\textmd{f}}$ in the disturbance-free case, and $u_{\textmd{b}}$ is the feedback controller to track the reference trajectory in the disturbance case. The feedforward controller is related to the initial state which is bounded in certain region, and the feedback controller is related to $x_{\textmd{f}}$ due to the tracking requirement. In the following, we first consider the nominal system of \eqref{eqn-4} to generate the feedforward controller and the reference trajectory in Section \ref{subsec-feedforward}, and then design the feedback controller for the system \eqref{eqn-4} in Section \ref{subsec-feedback}. Since the global STL specification has been decomposed into finite local ones in Section \ref{subsec-decomtime}, without loss of generality, we only focus on certain cell $\mathbf{Z}^{\ast}_{i}\in\mathbf{Z}^{\ast}$ with its time constraint $[0, \mathfrak{t}_{i}]$.

\subsection{Feedforward Controller Design}
\label{subsec-feedforward}

To design the feedforward controller, we consider the nominal system, which is the disturbance-free version of \eqref{eqn-4}:
\begin{equation}
\label{eqn-10}
\dot{x}(t)=Ax(t)+Bu(t).
\end{equation}
Let $\mathbf{x}_{\textrm{f}}(t, x_{0}, u)$ denote the solution of \eqref{eqn-10} at $t\in\mathbb{R}^{+}$ when the initial state is $x_{0}\in\mathbb{X}_{0}\subset\mathbb{X}$ and the control input is $u\in\mathbb{U}$.

For each $\mathbf{Z}^{\ast}_{i}\in\mathbf{Z}^{\ast}$, we denote it by $\mathbf{Z}^{\ast}_{i}=\{\mathbf{c}_{i}, \mathbf{G}_{i}, \mathbf{A}_{i}, \mathbf{b}_{i}\}$. From Section \ref{subsec-localSTL}, each $\mathbf{Z}^{\ast}_{i}$ has its own initial state set $\mathbf{I}^{\ast}_{i}:=\mathbf{Z}^{\ast}_{i}\cap\mathbf{Z}^{\ast}_{i-1}$. In particular, $\mathbf{I}^{\ast}_{0}=\mathbb{X}_{0}\cap\mathbf{Z}^{\ast}_{0}$. Since it is impossible to know which state to be chosen from $\mathbf{I}^{\ast}_{i}$ as the initial state, we need to consider the set $\mathbf{I}^{\ast}_{i}$ directly. Since $\mathbf{I}^{\ast}_{i}$ is a constrained zonotope, we denote it by $\{\bar{\mathbf{c}}_{i}, \bar{\mathbf{G}}_{i}, \bar{\mathbf{A}}_{i}, \bar{\mathbf{b}}_{i}\}$. Therefore, the feedforward controller is such that for each state $x_{0}\in\mathbf{I}^{\ast}_{i}$, the system state is driven into the target region $\mathbf{T}^{\ast}_{i}$. Here, we expect the final reachable set to be as close as possible to the center of $\mathbf{T}^{\ast}_{i}$ (denoted by $\mathbf{c}(\mathbf{T}^{\ast}_{i})$), which is formulated into the following optimization problem.
\begin{subequations}
\label{eqn-11}
\begin{align}
\label{eqn-11a}
\min_{u_{\textrm{f}}}\max_{x_{0}\in\mathbf{I}^{\ast}_{i}}&\  \|\mathbf{x}_{\textrm{f}}(\mathfrak{t}_{i}, x_{0}, u_{\textrm{f}}(\mathfrak{t}_{i}, x_{0}))-\mathbf{c}(\mathbf{T}^{\ast}_{i})\|+\int^{\mathfrak{t}_{i}}_{0}\|u_{\textrm{f}}(t, x_{0})\|dt \\
\text{s.t.}
\label{eqn-11c}
&\quad   u_{\textrm{f}}(t, x_{0})\in\mathbb{U}, \quad x_{0}\in\mathbf{I}^{\ast}_{i}, \quad t\in[0, \mathfrak{t}_{i}], \\
\label{eqn-11d}
&\quad   \mathbf{x}_{\textrm{f}}(t, x_{0}, u_{\textrm{f}}(t, x_{0}))\models\bar{\phi}_{i},   \\
\label{eqn-11e}
&\quad   \|\mathbf{G}^{+}_{i}(\mathbf{x}_{\textrm{f}}(t, x_{0}, u_{\textrm{f}}(t, x_{0}))-\mathbf{c}_{i})\|_{\infty}\leq1, \\
\label{eqn-11f}
&\quad    \mathbf{A}_{i}\mathbf{G}^{+}_{i}(\mathbf{x}_{\textrm{f}}(t, x_{0}, u_{\textrm{f}}(t, x_{0}))-\mathbf{c}_{i})=\mathbf{b}_{i},
\end{align}
\end{subequations}
where $\mathbf{x}_{\textrm{f}}(t, x_{0}, u_{\textrm{f}}(t, x_{0}))$ is the feedforward trajectory, $\mathbf{c}(\mathbf{T}^{\ast}_{i})$ is the center of $\mathbf{T}^{\ast}_{i}$, and $\mathbf{G}^{+}_{i}$ is the Moore-Penrose inverse of the matrix $\mathbf{G}_{i}$. The cost function in \eqref{eqn-11a} includes the difference between the final state and the desired target state as well as the reference input to be designed. In \eqref{eqn-11}, \eqref{eqn-11c} includes the input, state and time constraints; \eqref{eqn-11d} is to satisfy the local STL formula $\bar{\phi}_{i}$; \eqref{eqn-11e}-\eqref{eqn-11f} are the constraints from $\mathbf{Z}^{\ast}_{i}$. If $\mathbf{Z}^{\ast}_{i}$ is a zonotope, then \eqref{eqn-11f} is not needed.

To solve the problem \eqref{eqn-11}, we take advantage of the property of the (constrained) zonotope $\mathbf{I}^{\ast}_{i}$ denoted by $\{\bar{\mathbf{c}}_{i}, \bar{\mathbf{G}}_{i}, \bar{\mathbf{A}}_{i}, \bar{\mathbf{b}}_{i}\}$, and design the feedforward controller as follows:
\begin{align}
\label{eqn-12}
u_{\textrm{f}}(t, x_{0})=u_{\textrm{f}}(t, \bar{\mathbf{c}}_{i})+\sum^{p_{i}}_{l=1}\xi^{l}_{i}u_{\textrm{f}}(t, \bar{\mathbf{g}}^{l}_{i}),
\end{align}
where $x_{0}\in\mathbf{I}^{\ast}_{i}$, $\bar{\mathbf{G}}_{i}=(\bar{\mathbf{g}}^{1}_{i}, \ldots, \bar{\mathbf{g}}^{p_{i}}_{i})$ and $\xi_{i}=(\xi^{1}_{i}, \ldots, \xi^{p_{i}}_{i})$ with $\|\xi_{i}\|_{\infty}\leq1$ and $\bar{\mathbf{A}}_{i}\xi_{i}=\bar{\mathbf{b}}_{i}$. From \eqref{eqn-12}, the controller $u_{\textrm{f}}(t, x_{0})$ with $x_{0}\in\mathbf{I}^{\ast}_{i}$ depends on all input trajectories related to the center $\bar{\mathbf{c}}_{i}$ and all generators in $\bar{\mathbf{G}}_{i}$. Note that starting from any constrained zonotope, the reachable set has a center, which is expected to be as close as possible to $\mathbf{c}(\mathbf{T}^{\ast}_{i})$, and generators, which are expected to be as small as possible. Following this idea, we first design a reference controller starting from the center $\mathbf{c}_{i}$, and then design the feedforward controller starting from the set $\mathbf{I}^{\ast}_{i}$.

Starting from the center $\bar{\mathbf{c}}_{i}$, the design of the reference controller is summarized as the following optimization problem:
\begin{subequations}
\label{eqn-13}
\begin{align}
\label{eqn-13a}
\min_{u_{\textrm{f}}}&\  \|\mathbf{x}_{\textrm{f}}(\mathfrak{t}_{i}, \bar{\mathbf{c}}_{i}, u_{\textrm{f}}(\mathfrak{t}_{i}, \bar{\mathbf{c}}_{i}))-\mathbf{c}(\mathbf{T}^{\ast}_{i})\|+\int^{\mathfrak{t}_{i}}_{0}\|u_{\textrm{f}}(t, \bar{\mathbf{c}}_{i})\|dt \\
\text{s.t.}
\label{eqn-13c}
&\quad  u_{\textrm{f}}(t, \bar{\mathbf{c}}_{i})\in\mathbb{U},  \quad  t\in[0, \mathfrak{t}_{i}],  \\
\label{eqn-13d}
&\quad \mathbf{x}_{\textrm{f}}(t, \bar{\mathbf{c}}_{i}, u_{\textrm{f}}(t, \bar{\mathbf{c}}_{i}))\models\bar{\phi}_{i},   \\
\label{eqn-13e}
& \quad  \|\mathbf{G}^{+}_{i}(\mathbf{x}_{\textrm{f}}(t, \bar{\mathbf{c}}_{i}, u_{\textrm{r}}(t, \bar{\mathbf{c}}_{i}))-\mathbf{c}_{i})\|_{\infty}\leq1, \\
\label{eqn-13f}
&\quad  \mathbf{A}_{i}\mathbf{G}^{+}_{i}(\mathbf{x}_{\textrm{f}}(t, \bar{\mathbf{c}}_{i}, u_{\textrm{f}}(t, \bar{\mathbf{c}}_{i}))-\mathbf{c}_{i})=\mathbf{b}_{i}.
\end{align}
\end{subequations}
Comparing \eqref{eqn-11} with \eqref{eqn-13}, the only difference lies in that the initial state set $\mathbf{I}^{\ast}_{i}$ is considered in \eqref{eqn-11}, whereas only the center $\bar{\mathbf{c}}_{i}$ is addressed in \eqref{eqn-13}. The optimization problem \eqref{eqn-13} can be solved efficiently via many numerical optimization algorithms involving the choice of each step. Hence, we divide the time interval $[0, \mathfrak{t}_{i}]$ into $M_{i}\in\mathbb{N}$ parts uniformly, and derive a sequence of discrete times $\{t^{i}_{0}, \ldots, t^{i}_{M_{i}}\}$ with $t^{i}_{0}=0$ and $t^{i}_{M_{i}}=\mathfrak{t}_{i}$. Assume that the controller is piecewise continuous. The following auxiliary notations are defined:
\begin{align}
\label{eqn-14}
\bar{A}&=A^{M_{i}}, \quad  \bar{B}=(A^{M_{i}-1}B, \cdots, AB, B), \\
\label{eqn-15}
u_{\textrm{f}}(v)&=(u_{\textrm{f}}(t^{i}_{0}, v), \ldots, u_{\textrm{f}}(t^{i}_{M_{i}-1}, v)).
\end{align}
At the last time step, we can compute the state of \eqref{eqn-10} as $\mathbf{x}_{\textrm{f}}(t^{i}_{M_{i}}, x_{0}, u_{\textrm{f}})=\bar{A}x_{0}+\bar{B}u_{\textrm{f}}(x_{0})$ with $x_{0}\in\mathbf{I}^{\ast}_{i}$. Hence, \eqref{eqn-11a} can be achieved via the following:
\begin{align*}
&\min_{u_{\textrm{f}}}\left\|\bar{A}\bar{\mathbf{c}}_{i}+\bar{B}u_{\textrm{f}}(\bar{\mathbf{c}}_{i})+\sum^{p_{i}}_{l=1}\xi^{l}_{i}(\bar{A}\bar{\mathbf{g}}^{l}_{i}
+\bar{B}u_{\textrm{f}}(\bar{\mathbf{g}}^{l}_{i}))-\mathbf{c}(\mathbf{T}^{\ast}_{i})\right\| \\
&\qquad  +\left\|u_{\textrm{f}}(\bar{\mathbf{c}}_{i})+\sum^{p_{i}}_{l=1}\xi^{l}_{i}u_{\textrm{f}}(\bar{\mathbf{g}}^{l}_{i})\right\|,
\end{align*}
which can be rewritten as $\min_{u_{\textrm{f}}}\|\sum^{p_{i}}_{l=1}(\bar{A}\bar{\mathbf{g}}^{l}_{i}+\bar{B}u_{\textrm{f}}(\bar{\mathbf{g}}^{l}_{i}))\|+\|\sum^{p_{i}}_{l=1}u_{\textrm{f}}(\bar{\mathbf{g}}^{l}_{i})\|$ due to \eqref{eqn-13}.

\begin{proposition}
For the system \eqref{eqn-10}, the upper bound of the cost function in \eqref{eqn-11a} can be minimized via the following optimization problem:
\begin{subequations}
\label{eqn-16}
\begin{align}
\min_{u_{\textrm{f}}} &\  \sum^{p_{i}}_{l=1}(\|\bar{A}\bar{\mathbf{g}}^{l}_{i}+\bar{B}u_{\textrm{f}}(\bar{\mathbf{g}}^{l}_{i})\|+\|u_{\textrm{f}}(\bar{\mathbf{g}}^{l}_{i})\|) \\
\text{s.t.}
&\quad  u_{\textrm{f}}(\bar{\mathbf{c}}_{i})+\sum^{p_{i}}_{l=1}\xi^{l}_{i}u_{\textrm{f}}(\bar{\mathbf{g}}^{l}_{i})\in\mathbb{U}^{M_{i}},  \\
&\quad  \bar{\phi}_{i} \text{ is satisfied for the system \eqref{eqn-10}}.
\end{align}
\end{subequations}
\end{proposition}

In the feedforward controller design, the discrete-time sequence is applied to facilitate the computation. In addition, if we take piecewise continuous inputs, then we check the state and input constraints only at discrete times. To investigate the state evolution between any two successive discrete times, we can tighten the state constraint to be a subset of $\mathbf{Z}^{\ast}_{i}$.

\subsection{Feedback Controller Design}
\label{subsec-feedback}

The feedback controller is designed to minimize the effects of the disturbance while guaranteeing the satisfaction of the STL formula for the system \eqref{eqn-4}. To this end, we apply a linear feedback controller with a time-varying controller gain, which is designed as the following form:
\begin{align}
\label{eqn-17}
u_{\textrm{b}}(t, x_{\textmd{f}}(t))=K(t)(x(t)-x_{\textrm{f}}(t)).
\end{align}
Combining the feedforward controller $u_{\textrm{f}}$ from \eqref{eqn-11} and the feedback controller \eqref{eqn-17}, we have the following optimization problem for the closed-loop system.
\begin{subequations}
\label{eqn-18}
\begin{align}
\min &\   \|\mathbf{x}(t, x_{0}, u, w)-\mathbf{c}(\mathbf{T}^{\ast}_{i})\|+\|u(t)\| \\
\text{s.t.}
&\quad   \mathbf{x}(t, x_{0}, u, w)\models\bar{\phi}_{i}, \quad u\in\mathbb{U}, \quad w\in\mathbb{W},   \\
&\quad  \mathcal{R}(t, \mathbf{I}^{\ast}_{i})\subseteq\mathbf{Z}^{\ast}_{i}, \quad   x_{0}\in\mathbf{I}^{\ast}_{i}, \quad  t\in[0, \mathfrak{t}_{i}].
\end{align}
\end{subequations}
The problem \eqref{eqn-18} can be solved using LQR theory and the matrices $Q, R$ can be assumed to be diagonal to ease the computation. With the obtained $Q, R$ and the controller $u$, we can compute the controller gain $K$ at each discrete time. Hence, the feedback controller is designed for the system \eqref{eqn-4}.

In the applied approach, the constraint satisfaction is embedded in the optimization problems. Since the controller is designed for each cell, the optimal controller is not global but local. Both offline and online complexity is mixed. The optimization problems \eqref{eqn-13} and \eqref{eqn-16} can be solved offline, the complexity is not critical and many algorithms with polynomial time complexity can be applied; see, e.g., \cite{Kochdumper2021aroc, Houska2011acado, Herceg2013multi}. For the feedback controller design, the optimization problem \eqref{eqn-18} needs to be solved online, and the complexity mainly comes from the computation of the reachable set and algebraic operations on all cells. Due to the use of the piecewise linear controller, the online complexity is reasonable \cite{Schurmann2020optimizing}.

\section{Numerical Example}
\label{sec-example}

\begin{figure}[!t]
\begin{center}
\begin{picture}(50, 85)
\put(-62, -15){\resizebox{50mm}{35mm}{\includegraphics[width=2.5in]{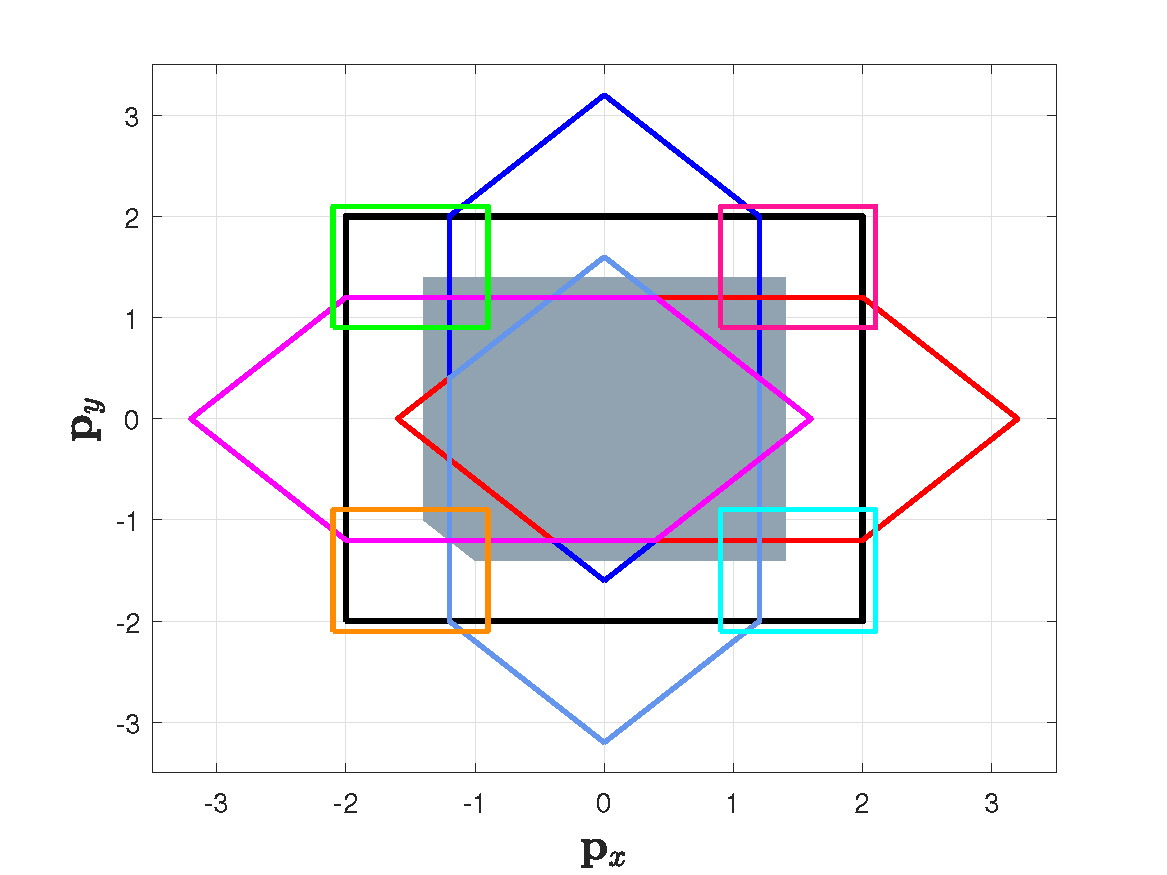}}}
\end{picture}\quad
\begin{picture}(50, 85)
\put(12, 0){\resizebox{35mm}{25mm}{\includegraphics[width=2.5in]{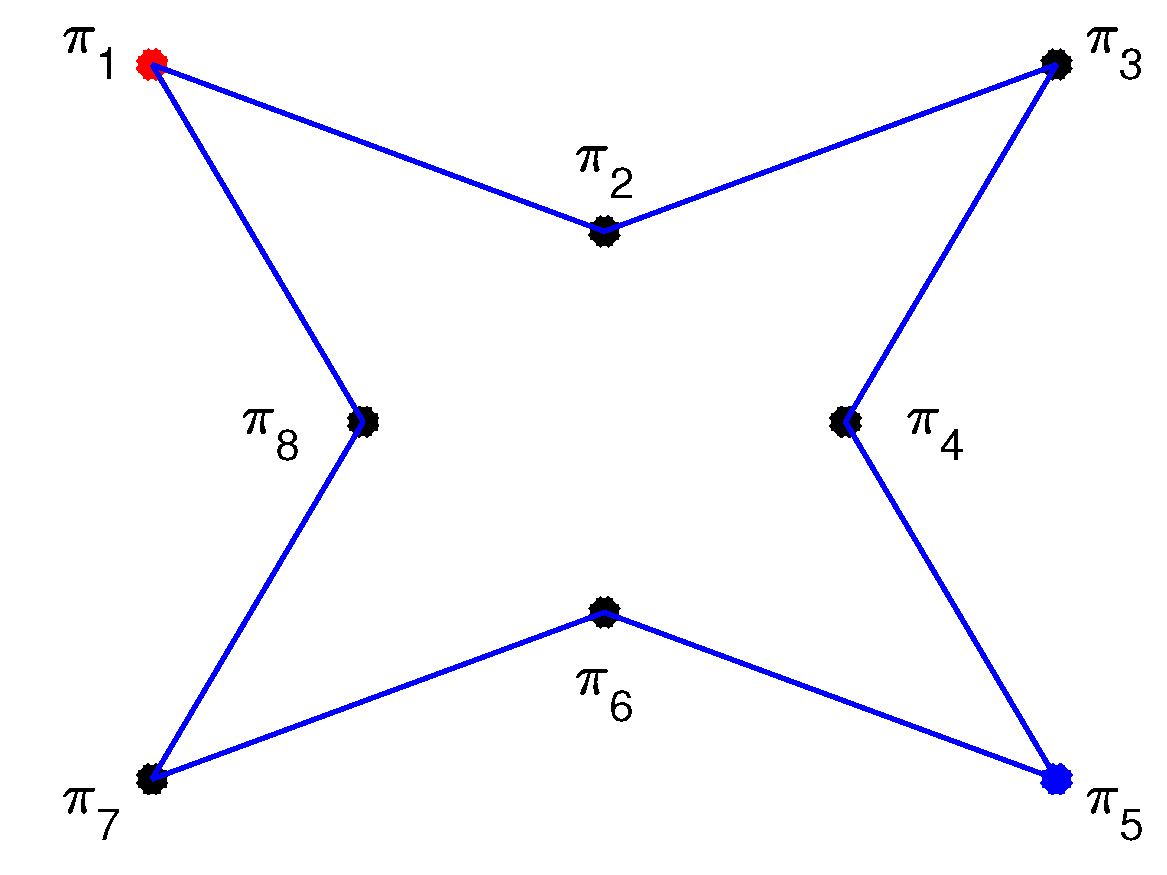}}}
\end{picture}
\end{center}
\caption{Illustration of the state space partition and the graph generation. Left: Both black and red lines are two paths for the LTL formula $\phi_{\textsf{LTL}}$.
Right: The graph generated from the intersection relation.}
\label{fig-2}
\end{figure}

In this section, a numerical example is presented to illustrate the derived results. All computations are performed on a laptop with Intel Core i7-10610U CPU@1.8GHz.

Consider an automated vehicle with the following double integrator dynamics:
\begin{align}
\label{eqn-19}
\dot{x}(t)=\begin{bmatrix}
0 & 1 & 0 & 0 \\
0 & 0 & 0 & 0 \\
0 & 1 & 0  & 0 \\
0 & 0 & 0 & 0
\end{bmatrix}x(t)+\begin{bmatrix}
0 & 0 \\
1 & 0  \\
0 & 0 \\
1 & 0
\end{bmatrix}u(t)+w(t),
\end{align}
where $x:=(\mathbf{p}_{x}, \mathbf{v}_{x}, \mathbf{p}_{y}, \mathbf{v}_{y})\in\mathbb{R}^{4}$ is the state, $u:=(\mathbf{a}_{x}, \mathbf{a}_{y})\in\mathbb{R}^{2}$ is the control input and $w\in\mathbb{R}^{4}$ is the external disturbance. In particular, $\mathbf{p}:=(\mathbf{p}_{x}, \mathbf{p}_{y})\in\mathbb{R}^{2}$ is the vehicle position, $\mathbf{v}:=(\mathbf{v}_{x}, \mathbf{v}_{y})\in\mathbb{R}^{2}$ is the vehicle velocity, and $(\mathbf{a}_{x}, \mathbf{a}_{y})\in\mathbb{R}^{2}$ is the acceleration. The state space is $\mathbb{X}:=[-2, 2]\times[-1, 1]\times[-2, 2]\times[-1, 1]$, the input set is $\mathbb{U}:=[-5, 5]\times[-5, 5]$, and the disturbance set is $\mathbb{W}:=[-0.05, 0.05]\times[-0.05, 0.05]\times[-0.05, 0.05]\times[-0.05, 0.05]$. The vehicle is to achieve the STL formula: $\phi:=\phi_{1}\wedge\phi_{2}$, where $\phi_{1}:=\textsf{G}_{[0, 7.5]}(x\in\mathbb{X}\wedge \mathbf{p}\notin\mathcal{O})$ and $\phi_{2}:=\textsf{F}_{[0, 7.5]}(\|\mathbf{p}-(1.7, -1.7)\|_{\infty}\leq0.2)$, which requires the vehicle to move to the red region shown in Fig. \ref{fig-2}. In particular, $\phi_{1}$ requires the vehicle to be in the state space while avoiding the obstacle denoted by $\mathcal{O}$ and shown in Fig. \ref{fig-2}. The initial state set is given as $\mathbb{X}_{0}=[-1.9, -1.5]\times[-1, 1]\times[1.5, 1.9]\times[-1, 1]$, which is the blue region in Fig. \ref{fig-2}.

To deal with this problem, we first partition the state space via the approach in Section \ref{sec-partition}. In particular, here we only partition the 2-D position space, and thus no velocity constraints are imposed in each cell. The partition is presented in Fig. \ref{fig-2} and results in 8 zonotopes, and thus no constrained zonotope is needed. From the intersection relation, the generated graph is given in Fig. \ref{fig-2}. For the STL-induced LTL formula $\phi_{\textsf{LTL}}$, we can find two admissible paths: $\mathbf{P}_{1}=\{\pi_{1}, \pi_{2}, \pi_{3}, \pi_{4}, \pi_{5}\}$ and $\mathbf{P}_{2}=\{\pi_{1}, \pi_{8}, \pi_{7}, \pi_{6}, \pi_{5}\}$. If we take the volume of cells as the evaluation function, then $\evl(\mathbf{P}_{1})=\evl(\mathbf{P}_{2})$, which indicates that either of these two paths can be chosen. If the volume of forbidden regions are taken into account, then we can see from \eqref{eqn-7} that the optimal path for $\phi_{\textsf{LTL}}$ is $\mathbf{P}_{2}$.

\begin{figure}[!t]
\begin{center}
\begin{picture}(60, 85)
\put(-60, -15){\resizebox{60mm}{35mm}{\includegraphics[width=2.5in]{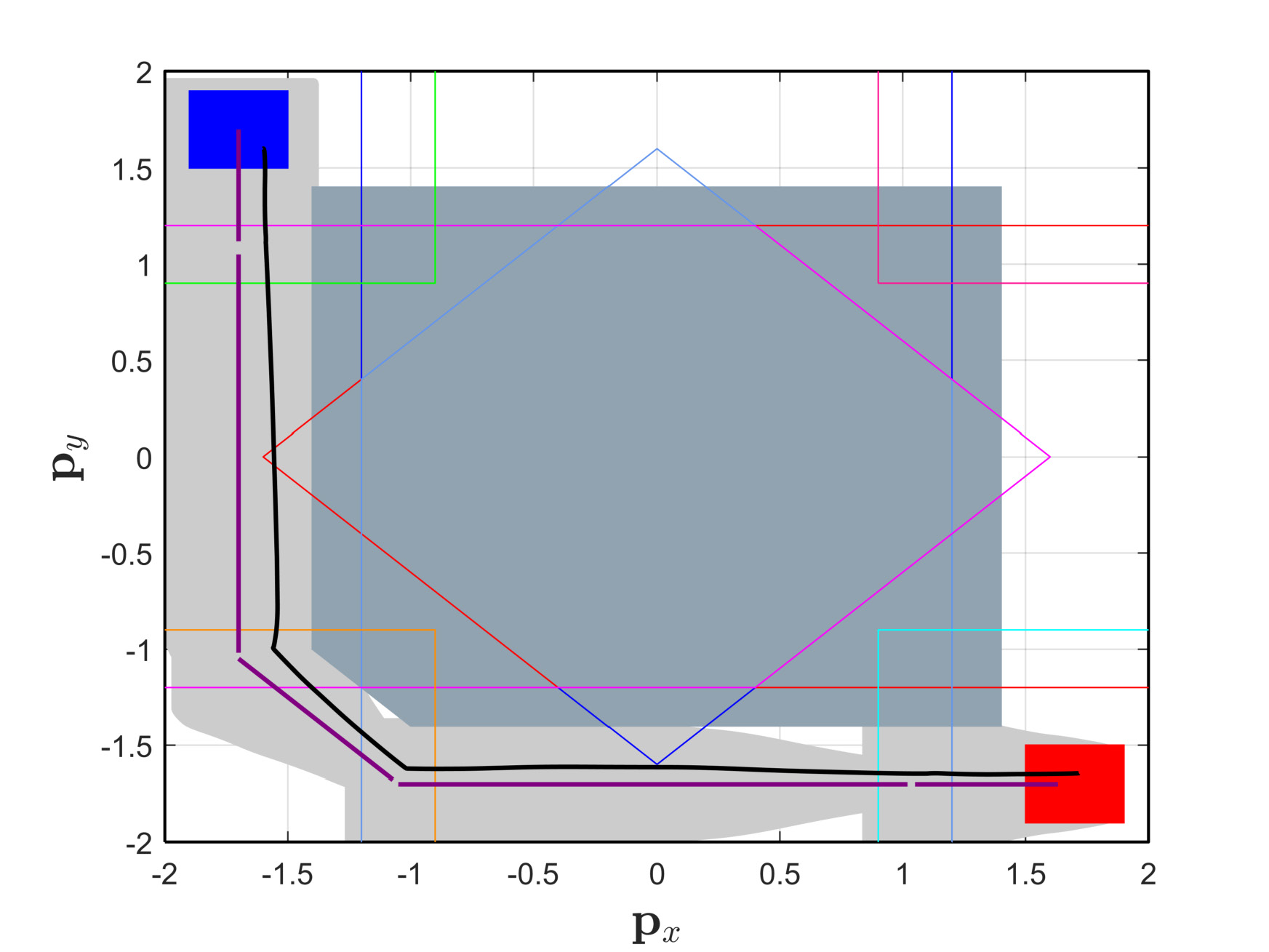}}}
\end{picture}
\end{center}
\caption{Trajectories for the vehicle system \eqref{eqn-19} with $x_{0}=(-1.6, 0, 1.6, 0)$ and $w=(0.05, 0.05, 0.05, 0.05)$. The blue region is the initial region, the red region is the goal region, the dark grey regions are obstacles, the purple lines are the reference trajectory, and the black curve region is the position trajectory of the vehicle.}
\label{fig-3}
\end{figure}

With the optimal path $\mathbf{P}^{\ast}=\mathbf{P}_{2}$, the time constraint is decomposed first, that is, $\mathfrak{t}_{1}=1, \mathfrak{t}_{2}=2, \mathfrak{t}_{3}=1.5, \mathfrak{t}_{4}=2$ and $\mathfrak{t}_{5}=1$. Next, the global STL formula $\phi$ is decomposed into five local STL formulas. For instance, in the region $\mathbf{Z}^{\ast}_{1}$, the local STL formula is given below.
\begin{align*}
\bar{\phi}_{1}&=\bar{\phi}_{11}\wedge\bar{\phi}_{12}\wedge\bar{\phi}_{13}, \quad \bar{\phi}_{11}=\textsf{G}_{[0, 7.5]}(x\in\mathbb{X}\wedge\mathbf{p}\notin\mathcal{O}), \\
\bar{\phi}_{12}&=\textsf{G}_{[0, 1]}(\mathbf{p}\in\mathbf{Z}_{1}),  \quad \bar{\phi}_{13}=\textsf{F}_{[0, 1]}(\mathbf{p}\in\mathbf{Z}_{1}\cap\mathbf{Z}_{8}).
\end{align*}
Similarly, we have the following local STL formulas for the regions $\mathbf{Z}^{\ast}_{2}, \mathbf{Z}^{\ast}_{3}, \mathbf{Z}^{\ast}_{4}, \mathbf{Z}^{\ast}_{5}$, respectively.
\begin{align*}
\bar{\phi}_{2}&=\bar{\phi}_{21}\wedge\bar{\phi}_{22}\wedge\bar{\phi}_{23}, \quad \bar{\phi}_{21}=\bar{\phi}_{11}, \\
\bar{\phi}_{22}&=\textsf{G}_{[1, 3]}(\mathbf{p}\in\mathbf{Z}_{8}), \quad \bar{\phi}_{23}=\textsf{F}_{[1, 3]}(\mathbf{p}\in\mathbf{Z}_{8}\cap\mathbf{Z}_{7}); \\
\bar{\phi}_{3}&=\bar{\phi}_{31}\wedge\bar{\phi}_{32}\wedge\bar{\phi}_{33}, \quad \bar{\phi}_{31}=\bar{\phi}_{11}, \\
\bar{\phi}_{32}&=\textsf{G}_{[3, 4.5]}(\mathbf{p}\in\mathbf{Z}_{7}), \quad \bar{\phi}_{33}=\textsf{F}_{[3, 4.5]}(\mathbf{p}\in\mathbf{Z}_{7}\cap\mathbf{Z}_{6}); \\
\bar{\phi}_{4}&=\bar{\phi}_{41}\wedge\bar{\phi}_{42}\wedge\bar{\phi}_{43}, \quad \bar{\phi}_{41}=\bar{\phi}_{11}, \\
\bar{\phi}_{42}&=\textsf{G}_{[4.5, 6.5]}(\mathbf{p}\in\mathbf{Z}_{6}), \quad \bar{\phi}_{43}=\textsf{F}_{[4.5, 6.5]}(\mathbf{p}\in\mathbf{Z}_{6}\cap\mathbf{Z}_{5}); \\
\bar{\phi}_{5}&=\bar{\phi}_{51}\wedge\bar{\phi}_{52}\wedge\bar{\phi}_{53}, \quad \bar{\phi}_{51}=\bar{\phi}_{11}, \\
\bar{\phi}_{52}&=\textsf{G}_{[6.5, 7.5]}(\mathbf{p}\in\mathbf{Z}_{5}), \\
\bar{\phi}_{53}&=\textsf{F}_{[6.5, 7.5]}(\|\mathbf{p}-(1.7, -1.7)\|_{\infty}\leq0.2).
\end{align*}
For each $i\in\{1, \ldots, 5\}$, $\bar{\phi}_{i1}$ corresponds to $\phi_{1}$ and is to ensure that the vehicle position is in the state space while avoiding the obstacle. $\bar{\phi}_{i2}$ and $\bar{\phi}_{i3}$ are the tasks for the region $\mathbf{Z}^{\ast}_{i}$.

\begin{figure}[!t]
\begin{center}
\begin{picture}(60, 80)
\put(-60, -16){\resizebox{60mm}{35mm}{\includegraphics[width=2.5in]{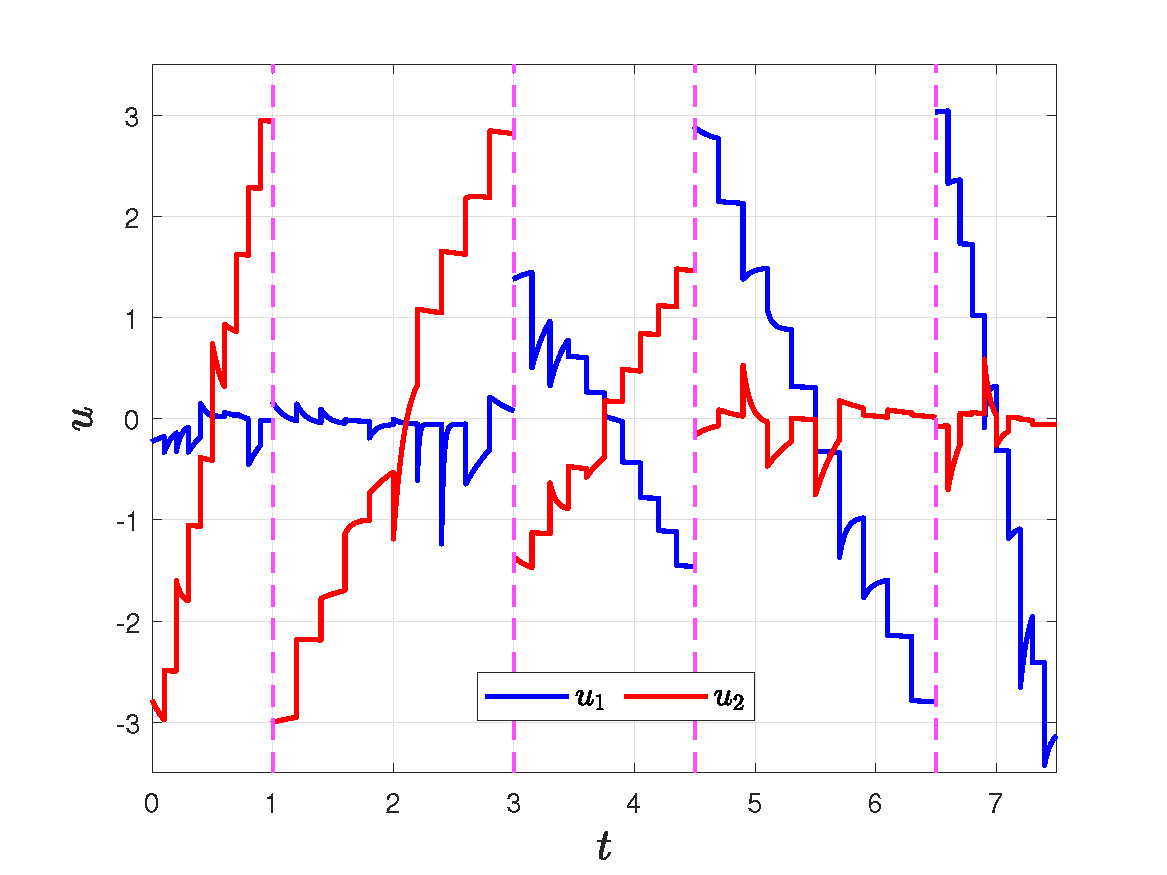}}}
\end{picture}
\end{center}
\caption{Control inputs for the vehicle system \eqref{eqn-19} with the STL formula $\phi$.}
\label{fig-4}
\end{figure}

In the following, we apply the proposed control strategy to design local controllers for the system \eqref{eqn-19} with all these local STL formulas. By solving the optimization problems in Section \ref{sec-optimalcontrol}, the position trajectories of the vehicle are presented in Fig. \ref{fig-3}, which shows the satisfaction of each local STL formulas and further the global STL formula $\phi$. To be specific, using AROC \cite{Kochdumper2021aroc}, we have all possible reachable sets, which are the gray region in Fig. \ref{fig-3}. By solving the optimization problem \eqref{eqn-11} via ACADO \cite{Houska2011acado}, the reference trajectory for the nominal system of \eqref{eqn-19} is derived as the purple lines (see Fig. \ref{fig-3}) for all regions $\mathbf{Z}^{\ast}_{i}$ with $i\in\{1, \ldots, 5\}$. Note that the gaps are caused by the minimization goal in \eqref{eqn-15} and many existing tools like MPT3.0 \cite{Herceg2013multi} can be implemented to generate the reference trajectory. Given the initial state $x_{0}=(-1.6, 0, 1.6, 0)$ and $w=(0.05, 0.05, 0.05, 0.05)$, the position trajectory of the vehicle is depicted as the black curve in Fig. \ref{fig-3}, and we can see that the external disturbance results in the difference between the reference trajectory and the simulated trajectory for the system \eqref{eqn-19}. The control inputs are given in Fig. \ref{fig-4}. We can see clearly from Fig. \ref{fig-4} that the control input experiences jumps when the local STL formulas switch. To show the satisfaction of the STL formula $\phi$ for all disturbances, we run 11 realizations of the disturbance trajectories and conclude that the STL formula $\phi$ is satisfied.

\section{Conclusion}
\label{sec-conclusion}

In this paper, we studied the controller synthesis problem for linear disturbed systems with STL specifications. We first combined the zonotope-based state space partition and the proposed evaluation mechanism to decompose the global STL formula into finite local ones. Second, based on reachability analysis and the properties of zonotopes, an optimization based method was proposed to design the controller with both feedforward and feedback parts. Finally, we presented a numerical example to demonstrate the proposed control strategy. Future work will be directed to the application of the proposed approach to nonlinear control systems and multi-agent systems with general STL specifications.


\end{document}